\documentclass[11pt,a4paper]{article}
\pdfoutput=1
\usepackage{amssymb}
\usepackage{amsmath}
\usepackage{latexsym}
\usepackage{amsfonts}
\usepackage{color,amssymb,amsmath,hyperref}
\usepackage{cite}

\newcommand{\de}{\partial} 
\def\be{\begin{equation}}
\def\ee{\end{equation}}
\def\bea{\begin{eqnarray}}
\def\eea{\end{eqnarray}}
\def\ba{\begin{array}}
\def\ea{\end{array}}
\def\e{\epsilon}
\def\wt{\widetilde}
\def\X{{\bf X}}
\def\K{{\bf K}}
\def\P{{\bf P}}

\def\q{{\bf q}}
\def\Q{\widehat{\bf Q}}

\def\D{\widehat{\bf D}}

\mathchardef\mhyphen="2D

\numberwithin{equation}{section}

\begin{document}

\begin{center}
\thispagestyle{empty}
{\huge  Bosonic Higher Spin Gravity 
in any Dimension}\\[5pt]
{\huge with Dynamical Two-Form}

\vskip 1cm

{\sc Cesar Arias}${}^{a,}$\footnote{ \href{mailto:cesar.arias@unab.cl}{\texttt{cesar.arias@unab.cl}}}, 
{\sc Roberto Bonezzi}${}^{b,}$\footnote{
\href{mailto:roberto.bonezzi@umons.ac.be}{\texttt{roberto.bonezzi@umons.ac.be}}}
{and \sc Per Sundell}${}^{a,}$\footnote{
\href{mailto:per.sundell@unab.cl}{\texttt{per.sundell@unab.cl}
}}\\[3pt]

\vskip .5cm

${}^a$\textit{Departamento de Ciencias Fisicas, Universidad Andres Bello}\\ 
\textit{Sazie 2212, Santiago de Chile}\\[3pt]

${}^b$\textit{Groupe de M\'ecanique et Gravitation}\\
\textit{Unit of Theoretical and Mathematical Physics}\\
\textit{University of Mons -- UMONS, 20 place du Parc, 7000 Mons, Belgium}\\[5pt]

\end{center}
\vskip 1cm
\paragraph{Abstract:} We first propose an alternative 
to Vasiliev's bosonic higher spin gravities
in any dimension by factoring out a modified $sp(2)$ 
gauge algebra.
We evidence perturbative equivalence of the two models, 
which have the same spectrum of Fronsdal fields at the
linearized level.
We then embed the new model into a flat Quillen superconnection 
containing two extra master 
fields in form degrees one and two; more generally, the superconnection
contains additional degrees of freedom associated
to various deformations of the underlying non-commutative
geometry.
Finally, we propose that by introducing first-quantized $sp(2)$ ghosts
and duality extending the field content, the Quillen flatness 
condition can be unified with the $sp(2)$ gauge conditions 
into a single flatness condition that is variational with 
a Frobenius--Chern--Simons action functional.

\vskip 8cm

\newpage

\clearpage
\setcounter{page}{1}

\tableofcontents

\section{Introduction}

Higher spin gravity concerns the extension of ordinary gravity 
by Fronsdal fields so as to facilitate the gauging of nonabelian 
higher spin symmetries.
Fully nonlinear higher spin gravities have been formulated by
Vasiliev by extending spacetime by internal non-commutative 
directions so as to obtain non-commutative geometries described
by Cartan integrable systems, first in four and lower spacetime dimensions \cite{Vasiliev:1990en,Vasiliev:1992av,Vas3D} by means of twistor 
oscillators, and later in arbitrary spacetime dimensions
\cite{Vasiliev:2003ev} using vector oscillators (for reviews, see 
\cite{Vasiliev:1999ba,Bekaert:2005vh,Didenko:2014dwa}).
The latter family is a direct generalization\footnote{Strictly speaking, the equivalence between the twistor and the vector formulations in four dimensions has been established only at the linearized level.} to any dimension
of the four-dimensional Type A model \cite{Sezgin:2003pt}, 
which consists perturbatively of one real Fronsdal field 
for every even spin, including a parity even scalar field.

In this paper, we revisit the family of Type A models in any dimension, 
first by modifying their internal $sp(2)$ gauging without affecting the 
higher spin gauge algebra nor the perturbative spectrum, and then 
by modifying the field content and the higher spin algebra.
The first step yields a model that agrees with Vasiliev's original 
model at the linearized level, and we shall argue that the two models
are perturbatively equivalent.
The latter step yields a distinct model with bi-fundamental higher
spin representations containing additional propagating degrees of freedom, 
which is a natural generalization of the four-dimensional 
Frobenius--Chern-Simons model proposed in \cite{Boulanger:2015kfa},
motivated primarily by the fact that the extended symmetries restrict
drastically the class of higher spin invariants, hence the
form of a possible effective action, thus improving upon the
predictive powers.

The modification is also motivated by the fact that it 
facilitates an off-shell formulation as a topological field 
theory directly in terms of differential forms on an extended 
non-commutative manifold with boundaries containing spacetime 
manifolds.
This formulation is akin to topological open string field 
theory \cite{Witten,Bershadsky,Gaberdiel:1997ia}, which
we consider to be a desirable feature in view of 
past Vasiliev inspired works \cite{Engquist:2005yt} (see also also \cite{Arias:2015wha,Bonezzi:2015lfa,Arias:2016agc})
on the tensionless limit of string theory in anti-de Sitter 
(for related holography motivated works, see \cite{Sundborg:2001,Sezgin:2002rt,Klebanov:2002ja,Engquist:2005yt})
as well as the more recent progress \cite{Boulanger:2015ova,Sleight:2017pcz,Vasiliev:2017cae} on relating the Fronsdal program \cite{Fradkin:1986qy} 
(for a review see \cite{Bekaert:2010hw}) to Vasiliev's 
formulation.

The perturbative spectrum of the Type A model on five-dimensional
anti de Sitter spacetime can be obtained by truncating the
supermultiplets of the first Regge trajectory of the Type 
IIB superstring on its maximally symmetric anti-de Sitter vacuum 
down to the maximal spin field in each supermultiplet,
save the two scalar fields of the Konishi multiplet.
The Type A models have also been proposed \cite{Engquist:2005yt} as bosonic 
truncations of effective descriptions of tensionless strings and 
membranes on anti-de Sitter backgrounds, as supported by various 
considerations based on holography \cite{Sundborg:2001,Sezgin:2002rt,Klebanov:2002ja},
whereby the natural candidates for holographic duals are free 
conformal field theories.
Thus, the Type A models may open up a new window to holography
permitting access to a wide range of physically interesting 
quantum field theories in four and higher dimensions, including 
four-dimensional pure Yang--Mills theories.

The symmetries of Vasiliev's equations, which one may
characterize as being star product local on the higher
dimensional non-commutative geometries, induce highly 
non-local symmetries of the effective deformed Fronsdal 
theory, causing a tension with the standard Noether procedure,
used as a tool for obtaining a classical action serving 
as a path integral measure, as substantiated by the results 
of \cite{Sleight:2017pcz}.
This fact, when taken together with the nature of the 
holographic duals and inspired by the on-shell approach 
to scattering amplitudes and topological field theory methods, 
suggests that the intrinsic spacetime formulation of higher 
spin dynamics as a stand-alone deformed Fronsdal theory 
without any reference to higher dimensional non-commutative 
geometries, is to be treated as a quantum effective theory 
without any classical limit, governed by higher spin gauge 
symmetry and unitarity. 
Accordingly, Vasiliev's equations, once subjected to proper 
boundary conditions on the extended non-commutative spaces
where they are formulated, should be equivalent to
quantum effective equations of motion in spacetime
for deformed Fronsdal fields.

As for the path integral formulation of higher spin gravity, it has 
thus been proposed \cite{Boulanger:2011dd,Boulanger:2012bj} (see also
\cite{Boulanger:2015kfa,Bonezzi:2016ttk} and the review \cite{Arias:2016ajh}) to use the 
language of topological quantum field theories on 
(higher dimensional) non-commutative Poisson manifolds, 
which naturally describes the Vasiliev's equations,
and provides the aforementioned link to underlying first-quantized 
topological field theories in two dimensions
\cite{Arias:2015wha,Bonezzi:2015lfa,Arias:2016agc}.
Thus, the basic rules for constructing the classical action are to
work with the basic $n$-ary products and trace operations
for non-commutative differential graded algebras, resulting 
in the notion of star-product local non-commutative topological field theories.
These theories have been proposed \cite{Boulanger:2015kfa} 
to admit boundary states weighted by boundary observables fixed 
essentially by the requirements of higher spin symmetry 
and admissibility as off-shell deformations of 
Batalin--Vilkovisky master actions; the simplest example 
of such deformations are off-shell topological invariants,
given by generalized Chern classes.
Of the latter, a subset do not receive any quantum corrections,
mainly due to the conservation of form degrees at the vertices,
and they reduce on-shell to classical higher spin invariants 
that one may propose are equal, once proper boundary conditions 
are imposed, to the free energy functionals of deformed Fronsdal 
theories; these ideas are substantiated by properties of
higher spin invariants closely related to the Chern classes,
known as zero-form charges 
\cite{Sezgin:2005pv,Colombo:2010fu,Sezgin:2011hq,Colombo:2012jx,Didenko:2012tv} 
(for recent progress, see \cite{Bonezzi:2017vha}).
The spectrum of boundary states and deformations is, however, 
much richer, and may hence open up new bridges between conformal 
and topological field theories; it would be interesting to
compare these to similar correspondences that have 
already been established using string and M--theory
\cite{Alday:2009aq,Book}.

In order to formulate Vasiliev's, or Vasiliev-like, higher spin gravities
as topological field theories, of key importance is the fact that the original 
Vasiliev system contains closed and central elements in form degree two, which
combine with the Weyl zero-form built on-shell from the Weyl tensors of the 
Fronsdal fields (and the scalar field), into deformations of the non-commutative 
structure on symplectic leafs of the base manifold.
Recently \cite{Boulanger:2015kfa}, the twistor formulation of four-dimensional
higher spin gravity has been modified such that the aforementioned 
closed and central elements arises as background values of a dynamical 
two-form master field, suggesting that the new theory possesses
a moduli space of non-commutative geometries.
A key feature of the new model is thus that it is formulated in terms of
only dynamical fields, which in the maximally duality extended case
form a gapless spectrum of forms, fitting into a Quillen superconnection
\cite{Quillen}, as would be expected from a theory
with a string-like first-quantized origin \cite{Witten,Bershadsky,Gaberdiel:1997ia}.
More precisely, the dynamical field content can be packaged into a
single Quillen superconnection \cite{Quillen} valued in a Frobenius algebra
akin to a topological open string field, leading to a renovated
version of the proposal of \cite{Engquist:2005yt}.
Indeed, a stringy feature of the model is that its moduli
contain various geometric deformations of the base manifold.
More precisely, some combinations of zero-form and two-form moduli deform
its symplectic structure, while others are transmitted into
Weyl tensors for Fronsdal fields.

A simple observation, which will be of importance in what follows,
is that the introduction of the dynamical two-form implies
that on a general background the equations of motion \emph{cannot}
be rewritten as a Wigner deformed oscillator algebra.
In the case of the four-dimensional twistor theory, this
implies that the Lorentz covariance can only be made
manifest within the Vasiliev-type phase, as here the
deformed oscillator algebra is restored.
As for the higher-dimensional vectorial models,
the consequences reach further, as the deformed
oscillator algebra enters the field dependent $sp(2)$
at the core of Vasiliev's original model.
In this paper, we shall instead factor out an $sp(2)$ algebra
with field independent generators, which we shall
refer to as $sp(2)^{(Y)}$, that does not refer
to any underlying Wigner deformed oscillator algebra.
At the linearized level, this implies that the classical
moduli appearing via vacuum expectation values of the
zero- and two-form consists of Fronsdal fields.

We would like to stress that the new model differs from Vasiliev's
original family of Type A models in two ways, as the latter
does not contain any dynamical two-form and
is based on representations obtained by
factoring out an $sp(2)$ algebra with
field dependent generators, constructed
using deformed Wigner oscillators as well as
undeformed oscillators, which we shall refer to
as $sp(2)^{(\rm diag)}$ as it is the manifest
$sp(2)$ symmetry acting by rotating all
doublet indices simultaneously.
However, despite this apparent advantage,
to our best understanding, the $sp(2)^{(\rm diag)}$ gauged
model does not admit any bi-fundamental extension nor can
it be coupled to a dynamical two-form.

We emphasize that the existence of two possible $sp(2)$ 
gaugings stems from the fact that both meet the basic criteria for
choosing the $sp(2)$ gauge algebra, namely Cartan
integrability of the full nonlinear system, and the
Central On Mass Shell Theorem \cite{Vasiliev:1999ba}, 
\emph{i.e.} consistency of the linearized system, 
as we shall spell out in detail in Section 3.
Thus, starting at the linearized level, where the two theories
are clearly equivalent, the old gauging is possible only on
special non-commutative base manifolds while the new gauging,
which is thus more akin to topological open string theory,
is distinguished by its potential extension to
general non-commutative base manifolds.

The paper is organized as follows: In Section 2, we review
selected features of Vasiliev's original formulation of higher
spin gravities in arbitrary dimensions.
In Section 3, we proceed with the formulation of the new model
based on a modified $sp(2)$ gauging. We compare the resulting 
new model with the original Vasiliev's Type A model at the (full) 
perturbative level as well as at the level of higher spin invariants, 
highlighting the crucial r\^ole played by the duality 
extension in the new model.
In Section 4, we couple the new model to a dynamical two-form
and further extend the system to a flat superconnection.  
Introducing $sp(2)$ ghosts we construct a BRST operator and 
propose an action principle that encodes the flatness condition
and $sp(2)$ invariance of the system.
We conclude in Section 5 pointing to a number of future directions.

\section{Vasiliev's Type A model}

In what follows, we outline Vasiliev's
original formulation of self-interacting totally symmetric
higher spin gauge fields in arbitrary spacetime dimensions.

\subsection{Master field equations}

A basic feature of Vasiliev's original theory, that will
remain essentially intact in the new theory, is the 
formulation of higher spin gravity in terms of horizontal
forms on non-commutative fibered spaces, which we refer to as
correspondence spaces.
The space of horizontal forms is a differential graded associatve
algebra, whose differential and binary product we shall denote by 
$d(\cdot)$ and $(\cdot)\star(\cdot)$, respectively.
Locally, these spaces are direct products of a base manifold with
coordinates $(X^M, Z^A_i)$ and line elements $(dX^M, dZ^A_i)$,
and a fiber space with coordinates $Y^A_i$.
The horizontal differential on the correspondence spaces is thus given by
\begin{equation}
d=dX^M\de_M+dZ^A_i\frac{\de}{\de Z^A_i}\;.
\end{equation}
Here $X^M$ coordinatize a commutative manifold, containing
spacetime, whereas $Z^A_i$ and $Y^A_i$ are non-commutative
coordinates, with non-trivial commutation relations
\begin{equation}\label{YZ commutators}
[Y^A_i,Y^B_j]_\star=2i\,\epsilon_{ij}\,\eta^{AB}\;,\qquad [Z^A_i,Z^B_j]_\star=-2i\,\epsilon_{ij}\,\eta^{AB}\;,
\end{equation}
where $\eta^{AB}$ is the $so(2,D-1)$ invariant symmetric tensor
and $\epsilon_{ij}$ is the $sp(2)$ invariant anti-symmetric tensor.
In order to define Lorentz tensors, one introduces a constant frame field
$(V_A,V_A^a)$ obeying $\eta^{AB} V_A V_B =-1$, $\eta^{AB}  V_A^a V_B=0$ and 
$\eta^{AB}  V_A^a V_B^b=\eta^{ab}$, and defines $Y_i:=V_A Y^A_i$ and 
$Y^a_i=V^a_A Y^A_i$ \emph{idem} $Z_i$ and $Z^a_i$.

The dynamical fields, all of which are horizontal, are a twisted-adjoint
zero-form $\Phi(X,Z;Y)$ and an adjoint one-form $W=dX^M\,W_M(X,Z;Y)+dZ^{Ai}\,W_{Ai}(X,Z;Y)$, 
which we shall refer to as master fields as they comprise infinite towers of tensor 
fields on the commuting manifold.
The system is put on-shell by imposing the constraints 
\begin{align}  
\label{Vasiliev1}
F+ \Phi \star J=0\ ,\qquad D \Phi&=0\ ,\\
\label{Vasiliev2}
D K_{ij}=0\ ,\qquad [K_{ij},\Phi]_\pi&=0\ ,
\end{align}
where $K_{ij}$ generate an $sp(2)$ algebra, 
\emph{viz.}
\be [K_{ij}, K_{kl}]_\star =4i \e_{(j|(k} K_{l)|i)}\ ,\label{fullsp2alg}\ee
which together form a quasi-free differential algebra, 
and factoring out the orbits generated by the shift symmetries
\be \delta W=K_{ij}\star \alpha^{ij}\;,\qquad 
\delta \Phi=K_{ij}\star\beta^{ij}\ ,\ee
where $\alpha^{ij}$ and $\beta^{ij}$ are triplets under the
adjoint and twisted-adjoint action of $sp(2)$,
respectively, \emph{viz.}
\begin{equation}
[K_{ij},\alpha^{kl}]_\star=4i\,\delta^{(k}_{(i}\alpha^{l)}_{j)}\;,\qquad
[K_{ij},\beta^{kl}]_\pi=4i\,\delta^{(k}_{(i}\beta^{l)}_{j)}\;.
\end{equation}
The equations of motion transform covariantly under gauge transformations
\be \delta_\e W=D\e\ ,\qquad \delta_\e\Phi=-[\e,\Phi]_\pi\ ,\qquad \delta_\e K_{ij}=-[\e,K_{ij}]_\star\ .\ee

In the above, the following definitions have been used:
The curvature and covariant derivatives
\begin{align}
F&:=dW+W\star W\;,\\
D\Phi&:=d\Phi+[W,\Phi]_\pi\;,\\
DK_{ij}&:=dK_{ij}+[W, K_{ij}]_\star\;,
\end{align}
where the $\pi$-twisted commutator
\begin{equation}\label{picomm}
[f,g]_\pi:=f\star g-(-1)^{{\rm deg}(f){\rm deg}(g)}g\star\pi(f)\ ,
\end{equation}
using the automorphism $\pi$ of the star product algebra defined by
\begin{equation}\label{pimap}
\pi(X^M,Z^a_i,Z_i;Y^a_i, Y_i):=(X^M,Z^a_i,-Z_i;Y^a_i, -Y_i)\;,\quad \pi d=d\pi\;.
\end{equation}
The element $J$ is a closed and central two-form
\be J=-\frac{i}4 dZ^i dZ_i ~\kappa \ ,\ee
where $\kappa$ is an inner Klein operator obeying
\be
dZ^idZ_i(\kappa\star f-\pi(f)\star\kappa)=0\;,\quad \kappa\star\kappa=1\;,
\ee
for general horizontal forms $f$.
It follows that

\begin{equation}\label{kappa}
\kappa=\kappa_Y\star\kappa_Z\;,
\end{equation}
where
\be
dZ^idZ_i(\kappa_Z\star f-\pi_Z(f)\star\kappa_Z)=0\;,\quad \kappa_Y\star f-\pi_Y(f)\star\kappa_Y=0
\ee
for general horizontal forms, and
\begin{equation}\label{pimapYZ}
\begin{split}
&\pi_Z(X^M,Z^a_i,Z_i;Y^a_i, Y_i):=(X^M,Z^a_i,-Z_i;Y^a_i, Y_i)\;,\quad \pi_Z d=d\pi_Z\;,\\[2mm]
&\pi_Y(X^M,Z^a_i,Z_i;Y^a_i, Y_i):=(X^M,Z^a_i,Z_i;Y^a_i, -Y_i)\;.
\end{split}
\end{equation}
Finally, the master fields obey the reality conditions 
\be W^\dagger=-W\;,\quad\Phi^\dagger=\pi(\Phi)\;,\quad J^\dagger=-J\;,\ee
where the hermitian conjugation operation is defined by
\be (df)^\dagger= d(f^\dagger)\ ,\qquad (f\star g)^\dagger=(-1)^{{\rm deg}(f){\rm deg}(g)}
g^\dagger\star f^\dagger\ ,\ee
\be (X^M,Y^A_i,Z^A_i)^\dagger=(X^M,Y^A_i,-Z^A_i)\ .\ee

\subsection{Diagonal $sp(2)$ generators}

In Vasiliev's Type A model, the $sp(2)$ gauge algebra is
taken to be generated by 
\be K_{ij}= K^{({\rm diag})}_{ij}:=K^{(0)}_{ij}-K^{(S)}_{ij}\ ,\qquad
K^{(0)}_{ij}:=K^{(Y)}_{ij}+K^{(Z)}_{ij}\ ,\label{gaugesp2}\ee
where the two first generators are field independent, \emph{viz.}
\be K^{(Y)}_{ij}:=\frac12 Y^A_{(i} \star Y_{j)A}\equiv K_{ij}\ ,\qquad K^{(Z)}_{ij}:=-\frac12 Z^A_{(i} \star Z_{j)A} \ ,\ee
and $K^{(S)}_{ij}$ is the field dependent generator
\be K^{(S)}_{ij}:=-\frac12 S^A_{(i} \star S_{j)A}\; ,\ee
built from the generalized Wigner deformed oscillator
\be S_{Ai}:=Z_{Ai}-2i W_{Ai}\ ,\qquad (S_{Ai})^\dagger=-S_{Ai}\ ,\ee
which is an adjoint element in the sense that
\be \delta_\e S_{Ai}=-[\e,S_{Ai}]_\star\ .\ee
The $sp(2)$ generators defined above form three copies of $sp(2)$, \emph{viz.} 
\begin{equation}
\begin{split}
[K^{(Y)}_{ij}, K^{(Y)}_{kl}]_\star &=4i \e_{(j|(k} K^{(Y)}_{l)|i)}\ ,\qquad 
[K^{(Z)}_{ij}, K^{(Z)}_{kl}]_\star =4i \e_{(j|(k} K^{(Z)}_{l)|i)}\ ,\\
[K^{(S)}_{ij}, K^{(S)}_{kl}]_\star &=4i \e_{(j|(k} K^{(S)}_{l)|i)}\ ,
\end{split}
\end{equation}
of which the latter follows from 
\begin{equation}
\begin{split}
&[S_{Ai},S_{Bj}]_\star =-2i\e_{ij}(\eta_{AB}-V_A V_B \Phi\star \kappa)\;,\\[2mm]
& S_{ai}\star \Phi- \Phi\star \pi(S_{ai})=0\ ,\quad S_i \star \Phi+\Phi\star \pi(S_i)= 0\ ,
\end{split}
\end{equation}
which is an equivalent way of writing $F_{Ai,Bj}=-\frac{i}2 \e_{ij} V_A V_B \Phi\star\kappa$ and $D_{Ai}\Phi=0$ as a direct sum of an undeformed 
oscillator $S_{ai}$ and a Wigner deformed oscillator $S_i:=V^A\,S_{Ai}$, 
with $\Phi$ playing the role of deformation parameter.

As for the $sp(2)$ invariance conditions, it follows from $D_M S_{Ai}=0$ and
$[S_{Ai},\Phi]_\pi=0$ that 
\bea \label{DK} 
D_M K^{({\rm diag})}_{ij}=0&\Leftrightarrow& [K^{(0)}_{ij},W_M]_\star =0\ ,\\{}
[K^{({\rm diag})}_{ij},\Phi]_\pi=0&\Leftrightarrow& [K^{(0)}_{ij},\Phi]_\star =0\ ,\eea
while
\be 
D_{Ai} K^{({\rm diag})}_{jk}=0\quad \Leftrightarrow \quad
[S_{Ai},K^{(0)}_{jk}-K^{(S)}_{jk}]_\star=0\quad\Leftrightarrow\quad
[K^{(0)}_{ij}, S_{Ak}]_\star =2i S_{A(i}\e_{j)k}\ ,\ee
from which it follows that
\begin{align} 
[K^{({\rm diag})}_{ij},K^{({\rm diag})}_{kl}]_\star &=4i \e_{(j|(k} \left(K^{(0)}_{l)|i)}+K^{(S)}_{l)|i)}\right)-[K^{(0)}_{ij},K^{(S)}_{kl}]_\star -[K^{(S)}_{ij},K^{(0)}_{kl}]_\star \cr
&= 4i \e_{(j|(k} \left(K^{(0)}_{l)|i)}-K^{(S)}_{l)|i)}\right)=4i \e_{(j|(k} K^{({\rm diag})}_{l)|i)}\ ,
\end{align}
\emph{i.e.} the desired $sp(2)$ commutation rules \eqref{fullsp2alg}.
Under a gauge transformation, one has
\be \delta_\e K^{({\rm diag})}_{ij}=-\delta_\e K^{(S)}_{ij}=-[\e,K^{(S)}_{ij}]\ ,\ee
and hence $\delta_\e K^{({\rm diag})}_{ij}=-[\e,K^{({\rm diag})}_{ij}]_\star$ holds true provided that 
\be[K^{(0)}_{ij},\e]_\star =0\ ,\ee
which is indeed compatible with \eqref{DK}.

\subsection{Symbol calculus, gauge conditions and $sp(2)$ symmetry}

Having specified the basic ingredients, the following observations
are in order:

Although there is no canonical way to realize the star product
as a convolution formula, there are two choices that are
particularly convenient for the most basic purposes.

As far as finding (perturbatively) exact solutions
is concerned, which shall be a topic below, it is convenient
to separate completely the $Y$ and $Z$ variables by representing
horizontal forms $f$ by their Weyl ordered symbols $f_{\rm W}=[f]_{\rm W}$,
where $[\cdot]_{\rm W}$ thus denotes the map sending an operator
to its Weyl ordered symbol, sometimes referred to as the Wigner map.
Conversely, we write $f=[f_{\rm W}]^{\rm W}$, where thus
$[\cdot]^{\rm W}$ is the inverse Wigner map sending
classical functions to operators.
One way of defining the  Wigner map, is to convert the operator
product $[f_W]^{\rm W}\star [g_W]^{\rm W}$ to a corresponding
non-local composition rule 
\be f_{\rm W}\star g_{\rm W} = [[f_W]^{\rm W}\star
[f_W]^{\rm W}]_{\rm W}\ ,\ee
for symbols, which  is given by the
twisted convolution formula
\begin{equation}\label{Weyl star}
\big(f_{\rm W}\star g_{\rm W}\big)(Y,Z)=\int d\mu\, d\tilde\mu\, e^{i( V^i_AU_i^A+\tilde V^i_A\tilde U_i^A)}f_{\rm W}(Y+U,Z+\tilde U)\,g_{\rm W}(Y+V, Z-\tilde V)\;,
\end{equation}
where $d\mu=(2\pi)^{-2(D+1)}d^{2(D+1)}Ud^{2(D+1)}V$, \emph{idem} $d\tilde\mu$.
It follows that
\be (f(Y)\star g(Z))_W=f_W(Y)\,g_W(Z)\ .\ee
In particular, in the case of the inner Klein operator \eqref{kappa},
one finds
\begin{equation}\label{kappa weyl}
\kappa_Y=\left[2\pi\delta^2(Y_i)\right]^{\rm W}\;,\quad \kappa_Z=\left[2\pi\delta^2(Z_i)\right]^{\rm W}\;,\quad \kappa=\left[(2\pi)^2\delta^2(Y_i)\delta^2(Z_i)\right]^{\rm W}\ .
\end{equation}

On the other hand, in order to describe asymptotically anti-de Sitter regions
using perturbatively defined Fronsdal tensors, one needs to use another 
ordering scheme in which all master fields are real analytic at $Y=0=Z$.
To this end, one may choose to work with normal ordered
symbols $f_{\rm N}=[f]_{\rm N}$ in terms of which the 
star product reads
\begin{equation}\label{Normal star}
\big(f_{\rm N}\star g_{\rm N}\big)(Y,Z)=\int d\mu\, e^{i V^i_AU_i^A}f_{\rm N}(Y+U,Z+U)\,g_{\rm N}(Y+V, Z-V)\;.
\end{equation}
Consequently,
\begin{equation}\label{kappa normal}
\kappa_Y=\left[2\pi\delta^2(Y_i)\right]^{\rm N}\;,\quad
\kappa_Z=\left[2\pi\delta^2(Z_i)\right]^{\rm N}\;,\quad\kappa=\left[\exp(iY^iZ_i)\right]^{\rm N}\;.
\end{equation}
It also follows that if $f=f(Y)$ and $g=g(Z)$ then
\be f_{\rm W}(Y)=f_{\rm N}(Y)\ ,\quad g_{\rm W}(Z)=g_{\rm N}(Z)\ .\ee
Working in normal order, one can show that \cite{Vasiliev:2003ev} the unfolded 
description of free Fronsdal fields, as spelled out by the Central On Mass
Shell Theorem \cite{Vasiliev:1999ba}, is contained in the equations
\begin{equation}
\left[F_{MN}\right]_{\rm N}\rvert_{Z=0}=0\;,\quad 
\left[D_M\Phi\right]_{\rm N}\rvert_{Z=0}=0
\end{equation}
in their free limit, obtained by expanding perturbatively around the
anti-de Sitter background
for $W$, provided that \emph{i}) all linearized symbols are real analytic
at $Y=0=Z$;  and \emph{ii}) the gauge condition 
\begin{equation}\label{axialgauge}
W_{ai}=0\;,\quad Z^i\left[W_i\right]_{\rm N}=0\;,
\end{equation}
which we shall refer to as the Vasiliev-Fronsdal gauge, holds in the 
linearized approximation.
More generally, we shall argue that in order to describe deformed Fronsdal 
fields in asymptotically anti-de Sitter spacetimes, conditions (\emph{i}) and (\emph{ii}) 
must be imposed in the leading order of the generalized Fefferman--Graham expansion 
to all orders in classical perturbation theory, together with boundary conditions 
at infinity of $Z$-space in addition, essentially as boundary conditions
on a gauge function and Weyl zero-form.

Turning to the $sp(2)$ gauging, the choice of $sp(2)$ generators
made in \eqref{gaugesp2} amounts to gauging the rigid transformations
that act by simultaneous rotation of the doublets $(Y^A_i,Z^A_i,
dZ^A_i,W_A^i)$, which is a manifest symmetry in normal order, 
due to the particular form of $\kappa_{\rm N}$ given in 
\eqref{kappa normal}.
This property of $sp(2)^{(\rm diag)}$ together with the fact that
its generators reduce to those of $sp(2)^{(Y)}$ in the free limit
was the rationale behind Vasiliev's original construction.
More precisely, factoring out $sp(2)^{(Y)}$ from the free theory yields
linearized fluctuations in $W_M$ and $\Phi$ consisting of unfolded 
Fronsdal tensors and corresponding Weyl tensors on-shell, respectively.

\section{New Type A model}
%
Examining Vasiliev's original formulation, one
notes that its consistency relies on the facts that
\begin{enumerate}
  \item[1)] The $sp(2)$ generators form a star product Lie algebra.
  \item[2)] The element $J$ is closed and central.
  \item[3)] The $sp(2)$ gauge conditions have the desired free limit (in perturbative expansion around the AdS vacuum).
\end{enumerate}
The key observation of this paper is that all of these conditions hold true as well
if one instead of $K^{({\rm diag})}_{ij}$ uses the undeformed $sp(2)$ generators\footnote{
The undeformed $sp(2)$ generators $K^{(0)}_{ij}$ or $K^{(Z)}_{ij}$ obey conditions
(1) and (2) but not (3).}
\be K_{ij}= K^{(Y)}_{ij}\;,\ee
which thus yields an alternative Type A model 
that is distinct from the original one, as we
shall demonstrate explicitly in the next section
by solving the two models perturbatively and
comparing the results.

Clearly, the two alternative Vasiliev-type models agree 
at the linearized level in a perturbative expansion around the 
standard anti-de Sitter vacuum, since $K_{ij}^{({\rm diag})}-K_{ij}$ 
are given by nonlinear corrections in such an expansion.

At the non-linear level, the key feature of the $sp(2)$ gauge
conditions is that the $sp(2)$ generators form an algebra, as this
assures that in applying classical perturbation theory to solve the
$Z$-space constraints there is no risk of encountering any
inconsistency in the form of additional algebraic constraints
in the remaining $X$-space constraints at $Z=0$.
In this sense, both $sp(2)^{({\rm diag})}$ and $sp(2)^{(Y)}$ gaugings
are admissible, even though the former is based on a symmetry 
that is manifest in any order (acting as rotations of the
doublets $(Y^A_i,Z^A_i,dZ^A_i,S^A_i)$), while the latter is
based on a symmetry that is manifest in Weyl order,and hence
in any ordering scheme related to Weyl order by means of 
re-orderings and gauge transformations\footnote{Formally,
a star product algebra is defined up to re-orderings generated
by totally symmetric poly-vector fields, which form symmetries
of trace operations given by integrals with suitable defined
measures; for details, see \cite{Kontsevich,Cattaneo}.}.

As we shall see below, for both models, the differential constraints
can formally be solved perturbatively for general zero-form initial 
data and gauge functions by working in a convenient gauge in Weyl order, 
that we shall refer to as the integrable gauge.
Based on existing results for similar perturbative expansions in the
four-dimensional twistor version of the Type A (and B) model, we
shall propose that for suitable initial data and gauge functions, 
the resulting field configurations can be mapped to the
Vasiliev--Fronsdal gauge (in which the normal ordered symbols
of the master fields have perturbative expansions in terms of
Fronsdal tensors that are weakly coupled at weak curvatures,
such as in asymptotically anti-de Sitter regions).

The aformentioned map is given by a similarity transformation that does 
not leave the $sp(2)^{(Y)}$ generators invariant.
Consequently, in the old model, the $sp(2)^{({\rm diag})}$ 
generators are field dependent in both the integrable and 
Vasiliev--Fronsdal gauges, while in the new model, the 
$sp(2)^{(Y)}$ gauge condition is imposed using field independent
generators in the integrable gauge and field dependent
similarity transformed $sp(2)$ generators in the Vasiliev--Fronsdal 
gauge.
Hence, strictly speaking, in the new model, we shall refer 
to \eqref{axialgauge} as the the Vasiliev--Fronsdal basis (rather
than gauge).

Below, we shall also propose to construct higher spin invariants, referred to
as zero-form charges \cite{Colombo:2012jx}, using trace operations
and quasi-projectors that annihilate the two-sided ideals
generated by the $sp(2)$-generators.
As the zero-form initial data in the integrable gauge
is related to that in the Vasiliev--Fronsdal gauge
by means of a nonlinear map, the zero-form charges have non-trivial 
perturbative expansions in the Vasiliev--Fronsdal gauge (which thus
provides observables in the asymptotic weak coupling region of spacetime 
\cite{Boulanger:2015kfa}).
Whether these two sets of observables can be used to map
the two type A models into each other remains an open
problem.

\subsection{Manifest $sp(2)^{(Y)} \times sp(2)^{(Z)}$ symmetry}

We would like to stress that the $sp(2)^{(Y)}$ transformations
can be made into a manifest symmetry of the equations of motion.
In fact, these equations can be rewritten as to exhibit an
even larger symmetry, generated by $sp(2)^{(Y)} \times sp(2)^{(Z)}$.
To this end, one first goes to Weyl order, in which the
symbol calculus takes the form
\be Y^A_i\star Y^B_j:=\left[Y^A_i Y^B_j\right]_{\rm W}+i\eta^{AB}\e_{ij}\ ,\qquad 
Y^A_i\star Z^B_j:=\left[Y^A_i Z^B_j\right]_{\rm W}\ ,\ee
\be Z^A_i\star Y^B_j:=\left[Z^A_i Y^B_j\right]_{\rm W}\ ,\qquad 
Z^A_i\star Z^B_j:=\left[Z^A_i Z^B_j\right]_{\rm W}-i\eta^{AB}\e_{ij}\ ,\ee
which indeed has manifest $sp(2)^{(Y)}\times sp(2)^{(Z)}$ symmetry.
Likewise, we recall that the inner Kleinian $\kappa$
can be rewritten as to make the $sp(2)^{(Y)}\times sp(2)^{(Z)}$
symmetry manifest, \emph{viz.}
\be  \kappa = \kappa_{Y} \star \kappa_{Z}\ ,\quad
\kappa_{Y} = \left[ 2\pi \delta^2(Y^i) \right]^{\rm W}\ ,\quad 
\kappa_{Z} = \left[ 2\pi \delta^2(Z^i) \right]^{\rm W}\ .\ee
Thus, in Weyl order, both the $\star$ product and the central element
$J$ are manifestly $sp(2)^{(Y)}\times sp(2)^{(Z)}$ invariant,
and hence they are in particular invariant under the
$sp(2)^{(Y)}$ symmetry used to gauge the new model.

\subsection{Perturbative solution in integrable gauge}

The differential equations in $X$-space can be solved using a gauge 
function, \emph{viz.}
\be W = L^{-1} \star (W'+d) \star L\ ,\qquad \Phi=L^{-1} \star  \Phi' \star \pi(L)\ ,\qquad W'_M=0\ .\ee
The primed fields, which are thus $X$-independent, obey the reduced equations
\be d' W' + W'\star W' + \Phi'\star J= 0\ ,\qquad  d' \Phi'+W'\star \Phi'-\Phi'\star \pi(W')=0\ ,\qquad d'=dZ^A_i \frac{\partial}{\partial Z^A_i}\ .\ee
Imposing an initial condition on the zero-form in Weyl order, \emph{viz.}
\be \left.\left[\Phi'\right]_{\rm W}\right|_{Z=0}=\left[C'\right]_{\rm W}\ ,\ee
and imposing the gauge condition
\be Z^i \left[W'_i\right]_{\rm W}=0\ ,\quad W'_{ai}=0\ ,\label{primegauge}\ee
the resulting solution space can be written as
\be \Phi'=C'\ ,\quad W' =  \sum_{n\geqslant 1}
w^{(n)} \star (C'\star\kappa_y)^{\star n}\ ,\quad \pi_Z(w^{(n)})=w^{(n)}\ ,
\label{primesolution}\ee
where the perturbative corrections can be grouped into a generating element
\be w' :=\sum_{n\geqslant 1} w^{(n)}\nu^n\ ,\quad w^{(n)}=dZ^i w^{(n)}_i(Z^j)\ ,\qquad \nu\in\mathbb C\ ,\ee
obeying the deformed oscillator problem \cite{Vas3D}
\be d'w' + w'\star w' + \nu j' =0\ ,\qquad j' := -\frac{i}4 dZ^i dZ_i \kappa_z\ .\ee
Its solutions\footnote{We expect the structure of the resulting
moduli space to resemble that of the four-dimnesional twistor 
formulation of the Type A model, which
decomposes into discrete branches, each labelled by a flat 
connection on $Z$-space, and coordinatized by (continuous)
zero-form initial data in their turn belonging to cells
separated by ``walls'' given by critical deformation
parameters; for details, see \cite{Sezgin:2005pv,Iazeolla:2011cb,Iazeolla:2007wt}.}   
can be obtained by adapting the method for the four-dimensional
twistor formulation of the Type A model spelled out in \cite{Sezgin:2005pv}, by 
introducing an auxiliary frame $U^\pm_i$ in $Z$-space defining
creation and annihilation operators $Z^\pm$, and 
representing the dependence of $w'_i$ on $Z^j$ as an inverse Laplace transform 
in the variable $Z^+ Z^-$, or equivalently, solving the problem using a basis
for symbols in $Z$-space defined using normal order, followed by mapping back
to Weyl order; for details on the latter approach, see \cite{Iazeolla:2011cb}.

We would like to note that so far we have not imposed any $sp(2)$ gauge 
conditions, and consequently we have treated the new and old models in parallel.

\subsection{Similarity transformation to Vasiliev--Fronsdal basis}

Let us proceed, still in parallel between the old and new models, by finding the
gauge function $L$ that brings the solution from the integrable gauge to the  
Vasiliev--Fronsdal basis obeying
\be Z^i [W_i]_{\rm N} = 0\ ,\label{zgauge1}\ee
where thus the gauge fields become Fronsdal tensors in weak coupling
regions.
To this end, it is useful to introduce the homotopy contractor
\be \rho_{\vec v}(f):= \left[\imath_{\vec v}({\cal L}_{v})^{-1} f_{\rm N}\right]^{\rm N}\ ,
\quad \vec v= Z^{Ai} \vec\partial^{(Z)}_{Ai}\ ,\ee
that can be used to invert the action of $d'$
on operators $f$ whose normal ordered symbols
obey $\imath_{\vec v} f_{\rm N}=0$,
\emph{viz.}
\be \rho_{\vec v} \,d' f=f-\delta_{0,{\rm deg}(f)} \left[f_{\rm N}|_{Z=0}\right]^{\rm N}\ .\ee
For explicit calculations, one can use
the integral representation
\be {\cal L}_{\vec v}^{-1}=\int_0^1 \frac{dt}t t^{{\cal L}_{\vec v}}\ ,\ee
which has a well-defined action on symbols defined in normal order 
that are real analytic in $Z$-space
at $Z=0$.
Thus $L$ can be obtained in normal ordered form by first expanding
\be L=\sum_{n\geqslant 0}L^{(n)}\ ,\label{L expansion}\ee
and then iterating \eqref{zgauge1}, 
which yields \cite{Colombo:2010fu}
\begin{equation}\label{L perturb}
\begin{split}
L^{(n)}&=-L^{(0)}\star\left(\sum_{n_1+n_2+n_3=n} \rho_{\vec v} \left((L^{-1})^{(n_1)}\star W^{\prime (n_2)}\star L^{(n_3)}\right)\right.\\[2mm]
&\hspace{15mm}+\left.\sum_{n_1+n_2=n} \rho_{\vec v} \left((L^{-1}-(L^{(0)})^{-1})^{(n_1)}\star d'  L^{(n_2)}\right)\right)\ ,
\end{split}
\end{equation}
for $n\geqslant 1$, as can be seen from
\be\label{rho W=0} 0=\rho_{\vec v}  W=\rho_{\vec v}\left( L^{-1}\star (W'+d')\star L\right)\ ,\ee
by using $d' L^{(0)}=0$ to write
\be L^{-1}\star d' L=d'\big((L^{(0)})^{-1}\star L\big)+\big(L^{-1}-(L^{(0)})^{-1}\big)\star d'L\ ,\ee
and $((L^{(0)})^{-1}\star L)|_{Z=0}=1$ to integrate
\be \rho_{\vec v} \, d_z((L^{(0)})^{-1}\star L)=(L^{(0)})^{-1}\star L-1\ .\ee
The relation now reads
\be
(L^{(0)})^{-1}\star L-1=-\rho_{\vec v}\big[L^{-1}\star W'\star L+\big(L^{-1}-(L^{(0)})^{-1}\big)\star d'L\big]\;,
\ee
and one recovers the perturbative solution \eqref{L perturb} by 
inserting the expansion \eqref{L expansion}, which is thus well-defined 
provided that the arguments of the homotopy contractors are real analytic 
in $Z$ space in normal order.

The latter problem is similar to that studied in the case of the four-dimensional 
twistor formulation of the Type A model, where it was found that $L^{(1)}$ exists
if the gauge function $L^{(0)}$ and the zero-form initial data $\Phi'$ are
Gaussian elements corresponding, respectively, to the anti-de Sitter vacuum
and fluctuations thereabout given by the particle and black-hole-like modes.
In what follows, we shall assume that an analogous result holds for the Type
A model in any dimension for $\Phi'$ consisting of particle modes, that is,
that it is possible to map initial data in lowest weight spaces to linearized
Fronsdal fields on-shell.

\subsection{$sp(2)$ gauging}

In order to gauge $sp(2)$, we first impose the $sp(2)$ invariance conditions, 
which we shall tend to next, after which we shall proceed by factoring out the 
corresponding ideals at the level of higher spin invariants.
As we shall see, the resulting $sp(2)$ gaugings are equivalent at the 
linearized level.

\subsubsection{$sp(2)$ invariance}

\paragraph{Old model ($sp(2)^{({\rm diag})}$).}

We recall that, in the old model, the $sp(2){({\rm diag})}$ invariance conditions read
\bea D_M K^{({\rm diag})}_{ij}=0& \Leftrightarrow & [K^{(Y)}_{ij}+K^{(Z)}_{ij},W_M]_\star =0\ ,
\\
D_{Ak}  K^{({\rm diag})}_{ij}=0&\Leftrightarrow & [K^{(Y)}_{ij}+K^{(Z)}_{ij},S_{Ak}]_\star= 4i\,S_{A(i}\e_{j)k}\ ,\\{}
 [\Phi,K^{({\rm diag})}_{ij}]_\star=0& \Leftrightarrow & [K^{(Y)}_{ij}+K^{(Z)}_{ij},\Phi]_\star =0\ .\eea
In the integrable gauge, these conditions are equivalent to
\be [K^{(Y)}_{ij},C']_\star=0\ .\ee
In the Vasiliev--Fronsdal gauge, the $sp(2)^{({\rm diag})}$ invariance holds provided that
\be [K_{ij}^{(Y)}+K_{ij}^{(Z)},L^{(0)}]_\star=[K^{(Y)}_{ij},L^{(0)}]_\star =0\ ,\ee
as this condition implies that $[K_{ij}^{(Y)}+K_{ij}^{(Z)},L]_\star=0$ by virtue of the fact that the homotopy contractor $\rho_{\vec v}$ is $sp(2)_{\rm diag}$ invariant.

\paragraph{New model ($sp(2)^{(Y)}$).}

In the integrable gauge, the $sp(2)^{(Y)}$ invariance conditions reads (${K_{ij}\equiv
K^{(Y)}_{ij}}$)
\be [K_{ij},W']_\star=0=[K_{ij},\Phi']_\star\ ,\ee
which are equivalent to
\be [K_{ij},C']_\star=0\ .\ee
In the Vasiliev--Fronsdal basis the fields obey
the following similarity transformed $sp(2)^{(Y)}$ invariance conditions:
\be [\Phi,K^{(L)}_{ij}]_\star=0\ ,\qquad
DK^{(L)}_{ij}\equiv dK^{(L)}_{ij}+[W,K^{(L)}_{ij}]_\star=0\ ,\ee
where
\be K^{(L)}_{ij}:=L^{-1}\star K_{ij}\star L=
((L^{(0)})^{-1}\star L)^{-1}\star K_{ij}\star (L^{(0)})^{-1}\star L\ ,\ee
which are field dependent generators such that $(K^{(L)}_{ij})^{(0)}=K_{ij}$.

\paragraph{Equivalence between old and new model.}

In the Vasiliev--Fronsdal gauge, and prior to factoring 
out the ideal, both models have perturbatively defined solution 
spaces obeying the same differential equations, gauge conditions,
\emph{viz.}
\be W_{ai}=0\ ,\qquad Z^i W_i=0\ ,\ ,\ee
and $sp(2)$ invariance conditions, \emph{viz.}
\be D_i K_{jk}=0\ ,\qquad [K_{ij},\Phi]_\pi=0\ ,\qquad [K_{ij},K_{kl}]_\star
=4i\e_{jk} K_{il}\ .\ee
with $sp(2)$ generators subject to the same functional initial condition,
\emph{viz.}
\be K_{ij}|_{\Phi=0}=K^{(Y)}_{ij}\ .\ee
This suggests that the two models are perturbatively equivalent,
modulo redefinitions of zero-form initial data and modifications
of the Vasiliev--Fronsdal gauge condition away from the asymptotic
region.
This could be examined by comparing the first order corrections to 
$K^{(L)}_{ij}$ and $K^{({\rm diag})}_{ij}$, which we leave for a 
separate work.

\subsubsection{Factoring out the $sp(2)$ ideal}

Thus, the perturbatively defined configurations \eqref{primesolution} 
with $sp(2)$-invariant zero-form initial data obey the differential 
equations of motion as well as the $sp(2)$ invariance conditions
in the old as well as new models.
In both models, the problem of factoring out the $sp(2)$ orbits from
these solution spaces combines 
naturally with the problem of constructing higher spin invariants.

The (two-sided) ideal ${\cal I}$ in the algebra ${\cal A}_0$ of $sp(2)$ 
invariant master fields generated by the $sp(2)$ gauge algebra can be 
factored out from invariants by using the trace operation
\be {\rm Tr}_M \, [f]:={\rm Tr}\,M\star f\ ,\ee
where $[f]\in {\cal A}_0/{\cal I}$ is the equivalence class of
$f\in{\cal A}_0$; ${\rm Tr}$ is the trace operation on $Y$- and $Z$-space,
and $M$ obeys
\be K_{ij}\star M=0=M\star K_{ij}\ ,\ee
the covariant constancy condition
\be D_M M=0\ ,\qquad [B,M]_\star =0\ ,\qquad B:=\Phi\star \kappa\ ;\ee
and is a quasi-projector in the sense that $M\star {\cal A}_0$ 
exists (but not $M\star M\star  {\cal A}_0$).
In the new model, we have \cite{Sagnotti:2005ns}
\be M=M^{(Y)}=F(K^{ij\,(Y)}K_{ij}^{(Y)})\ , \ee
where $F$ is real analytic and nonvanishing at the origin, and
\be M^{(L)}=L^{-1}\star M^{(Y)}\star L=M^{(Y)}+{\rm h.o.t.} ,\ee
in the Vasiliev--Fronsdal basis; in the old model, we have 
\be M=M^{({\rm diag})}=M^{(Y)}+{\rm h.o.t.}\ ,\ee
where the higher order terms can be found by solving 
$K^{({\rm diag})}_{ij}\star M^{({\rm diag})}=0$
perturbatively \cite{Sagnotti:2005ns}.
It follows that 
\be D_M (M\star B)= 0\ ,\qquad D_M (M\star S_{Ai})=0\ ,\ee
of which the first equation indeed contains the correct 
linearized mass-shell conditions for generalized Weyl 
tensors (including the dynamical scalar field) \cite{Sagnotti:2005ns}.

The simplest invariants are the zero-form charges 
\cite{Sezgin:2005pv,Sezgin:2011hq} given by
\be {\cal O}_C:={\rm Tr}_M \, {\cal W}_C(S)\ ,\ee
where ${\cal W}_C$ are twisted (open) Wilson lines along curves 
$C$ from $Z=0$ to $Z=\Lambda(C)$, which can be straightened
out into star products of vertex-like operators \cite{Gross:2000ba,Bonezzi:2017vha},
\emph{viz.}
\be {\cal W}_C=f_C(B)\star {\cal V}_\Lambda\ ,\qquad {\cal V}_\Lambda:=
\exp_\star(i\Lambda^{Ai} S_{Ai})\ ,\ee
where $f_C$ is a star function (\emph{i.e.} its dependence on
$B$ is in terms of monomials $B^{\star n}$ for $n=0,1,2,\dots$) 
depending on the shape of $C$.
The zero-form charges are de Rham closed by virtue of 
\be \partial_M {\cal O}={\rm Tr}\, D_M(M \star {\cal W}_C)=0\ ,\ee
and hence higher spin invariant.

More general invariants \cite{Vasiliev:1999ba,Sezgin:2011hq}, that can 
be evaluated on non-trivial elements
$[\Sigma]$ in the singular homology of $X$-space, can be constructed by 
choosing a structure group $G$ with connection $\Omega_M$ and splitting 
\be W_M = \Omega_M+E_M\ ,\ee
where $E_M$ is a soldering one-form, that is, a generalized frame field,
whose gauge parameters belong to sections that can be converted to 
globally defined vector fields on $X$ (modulo a $G$ gauge transformation
with composite parameter).
This faciliates the definition of $G$-invariant tensors on $X$-space,
which induce top forms on representatives $\Sigma'\in [\Sigma]$ whose
integrals over $\Sigma'$ define generalized volumes whose
extrema (as one varies $\Sigma'$) are diffeomorphism
invariants, and hence higher spin gauge invariant by the soldering
mechanism.
These geometries also support closed abelian even forms 
\be H_{[2p]}={\rm Tr}_M\, (E\star E)^{\star p}\ ,\ee
on $X$-space, whose charges $\oint_\Sigma H_{[2p]}$
are higher spin gauge invariant.

As first suggested in \cite{Engquist:2005yt}, 
the zero-form charges have perturbative
expansions over asymptotically anti-de Sitter
solutions in terms of boundary correlation function,
as has been verified and developed further in the 
context of four-dimensional twistor oscillator models
\cite{Colombo:2010fu,Colombo:2012jx,Didenko:2012tv},
where it has also been proposed \cite{Iazeolla:2011cb} 
that they can be interpreted as extensive charges for families of localizable black-hole like solutions.
Thus, zero-form charges together with other invariants could serve as tools for
establishing the perturbative equivalence
between the old and new Type A models\footnote{They 
could also be useful in establishing the equivalence
between the vector and twistor oscillator formulations
of theType A model in four dimensions.}.

\section{Coupling of the new Type A model to a dynamical two-form}

The new Type A model can be coupled to a dynamical two-form, 
leading to an extended higher spin gravity model of
Frobenius--Chern-Simons type
based on a superconnection suitable for an off-shell formulation
and possibly also for making contact with topological open strings.

\subsection{Master field equations}

We introduce two separate connections $A$ and $\wt A$, with curvatures
\be  F:=dA+A\star A\ ,\qquad \wt F:=d\wt A+\wt A\star \wt A\ ,\ee
and a two-form $\wt \Phi$, and take $(\Phi,\tilde\Phi)$
to transform in opposite twisted bi-fundamental representations,
with covariant derivatives
\be D\Phi:=d\Phi+A\star \Phi-\Phi\star \pi(\wt A)\ ,\qquad \wt D\wt \Phi:=d\wt \Phi+
\pi(\wt A)\star \wt \Phi-\wt \Phi\star  A\ ,\label{tildethree}\ee
such that $\Phi\star\wt \Phi$ and $\pi(\wt \Phi\star \Phi)$ can be used 
to source $F$ and $\wt F$, respectively.
The resulting Cartan integrable equations of motion read
\begin{align}
\label{tildeone}
F+ \Phi \star \wt \Phi&=0\ ,\qquad 0=\wt F+ \pi(\wt \Phi \star  \Phi)\ ,\\
\label{tildetwo}
D \Phi&=0\ ,\qquad 0=\wt D \wt \Phi\ ,\\
D K_{ij} &=0\ ,\qquad 0=\wt D K_{ij} \ ,\\
\label{gaugesp2new}
[K_{ij},\Phi]_\pi &=0\ ,\qquad 0=[K_{ij},\pi(\wt \Phi)]_\pi \ ,
\end{align}
where $K_{ij}$ form a star product $sp(2)$ algebra that reduce to $K^{(Y)}_{ij}$
in the free limit, and field configurations are considered to be equivalent 
if they belong to the same orbit generated by the shift symmetries
\begin{equation}\label{factorout}
\delta\wt \Phi=K_{ij}\star\tilde\beta^{ij}\;,\quad \delta\wt A=K_{ij}\star\tilde\alpha^{ij}\;,\quad
\delta A=K_{ij}\star\alpha^{ij}\;,\quad \delta \Phi=K_{ij}\star\beta^{ij}
\end{equation}
for general undeformed $sp(2)$-triplets $(\tilde \beta^{ij},\tilde \alpha^{ij},\alpha^{ij},\beta^{ij})$.
Finally, its reality conditions are
\be \label{tilde reality}
A^\dagger=-\wt A\;,\quad \Phi^\dagger=\pi(\Phi)\;,\quad \wt\Phi^\dagger=-\pi(\wt \Phi)\;.
\ee

The equations can be re-written by introducing an outer Klein operator 
$k$ that obeys $k^2=1$ along with
\be [k,Y^a_i]=0\ ,\quad \{k ,Y_i\}=0\; ,\quad  
[k,Z^a_i]=0 \ ,\quad \{k ,Z_i\}=0\ ,\quad dk=kd\;,\ee
and defining 
\begin{equation}\label{reduction to twist}
\quad B=\Phi\,k\;,\quad\wt B=k\,\wt\Phi\;,
\end{equation}
after which the equations read
\begin{align}
\label{tildeoneEXT}
F+ B \star \wt B&=0\ ,\qquad 0=\wt F+ \wt B \star B\ ,\\
\label{tildetwoEXT}
D B&=0\ ,\qquad 0=\wt D \wt B\ ,\\
D K_{ij} &=0\ ,\qquad 0=\wt D K_{ij} \ ,\\
\label{gaugesp2newEXT}
[K_{ij},B]_\star &=0\ ,\qquad 0=[K_{ij},\wt B]_\star \ ,
\end{align}
where now
\be DB:=dB+A\star B-B\star \wt A\ ,\qquad 
\wt D\wt B:=d\wt B+\wt A\star \wt B-\wt B\star  A\ ,
\label{tildethreeEXT}\ee
and the $sp(2)^{(Y)}$ gauge symmetries read
\begin{equation}\label{factoroutEXT}
\delta\wt  B=K_{ij}\star\tilde\beta^{ij}\;,
\quad \delta\wt A=K_{ij}\star\tilde\alpha^{ij}\;,\quad
\delta A=K_{ij}\star\alpha^{ij}\;,\quad \delta  B=K_{ij}\star\beta^{ij}
\end{equation}
for general undeformed $sp(2)^{(Y)}$-triplets 
$(\tilde \beta^{ij},\tilde \alpha^{ij},\alpha^{ij},\beta^{ij})$.
The reality conditions are
\be \label{tilde realityEXT}
A^\dagger=-\wt A\;,\quad B^\dagger=B\;,\quad \wt B^\dagger=-\wt B\;.
\ee

The system can be extended further in two independent ways, 
by allowing general dependence on $k$, and by duality extension, 
whereby $(A,\wt A,B,\wt B)$ are forms of degrees $(1,1,0,2)$ 
mod $2$, respectively.
Reducing the $k$-dependence by taking $B=\Phi k$ and 
$\widetilde B=k\widetilde \Phi$ and $(A,\widetilde A,\Phi,\widetilde \Phi)$ 
to be $k$-independent forms of degrees $(1,1,0,2)$ mod $2$, respectively, 
yields the duality extension of the original system with twisted 
bi-fundamental zero- and two-form. 

Prior to eliminating $k$, the one-form $S:=dZ^{Ai}S_{Ai}$ with
$S_{Ai}:=Z_{Ai}-2i A_{Ai}$ obeys
\begin{align}
\label{EXT non osc}
[S_{ai},S_{bj}]_\star&=2\,\imath_{bj}\imath_{ai}(S\star S)\;,\\
\pi_k(S_{Ai})\star S_j-S_j\star S_{Ai}&=2\,\imath_{j}\imath_{Ai}(S\star S)\;,
\end{align}

where 
\be
S\star S=i\,dZ^{Ai}dZ_{Ai}+4B\star\wt B\;,
\ee
and the inner derivatives $\imath_{Ai}\equiv\imath_{\de_{Ai}}$
act from the left, using the rule $[k,\imath_{ai}]=0$ and
$\{k,\imath_i\}=0$.
In deriving \eqref{EXT non osc} we have used 
$\{dZ^{Ai}Z_{Ai},A\}=-2idZ^{Ai}\de_{Ai}A$ and 
$F=-B\star\wt B$.
Thus, after eliminating $k$, we have
\be [S_{Ai},S_{Bj}]_\star=2\,\imath_{Bj}\imath_{Ai}(S\star S)\;,\quad
S\star S=i\,dZ^{Ai}dZ_{Ai}+4\Phi\star\wt \Phi\;,\ee
that is, the presence of the dynamical two-form implies that $S_{Ai}$
is no longer a deformed oscillator on-shell.
The one-form $\wt S:=dZ^{Ai}S_{Ai}$ with
$S_{Ai}:=Z_{Ai}-2i \widetilde A_{Ai}$ obeys similar constraints, 
and we note that there is no constraint on mutual star products 
between $S_{Ai}$ and $\wt S_{Ai}$ master fields.

As for the choice of $sp(2)$ gauge algebra generators, the introduction
of the dynamical two-form obstructs the Wigner deformed oscillator
algebra, and hence the definition of a diagonal $sp(2)$ algebra.
On the other hand, the choice 
\be K_{ij}=K^{(Y)}_{ij}\ ,\ee
remains consistent for general two-form backgrounds.
With this choice, and assuming that ${\cal Z}$ contains an 
$S^2$ on which $\widetilde B$ can be wrapped as to produce $J$ 
as a vacuum expectation value, the consistent truncation
\be\wt\Phi=J\;,\quad\wt A = A = W \;,\label{truncation to Vasiliev}\ee
gives back the new Type A model. 
The non-trivial two-cycle implies, however, that the dynamical 
two-form contains additional degrees of freedom, that we plan
to examine elsewhere; for a related feature in the case of
four-dimensional higher spin gravity, see \cite{Boulanger:2015kfa,Bonezzi:2016ttk}.

\subsection{Frobenius algebra and superconnection}

As topological open strings set the paradigm for deforming 
differential form algebras on Poisson manifolds
\cite{Arias:2015wha,Bonezzi:2015lfa,Arias:2016agc,Kontsevich,Cattaneo,Chu,Beggs,Zumino},
this raises the question of whether the field equations 
admit a format more akin to that expected from a topological 
open string field theory, namely that of a flatness condition 
on a graded odd superconnection valued in the
direct product of the higher spin algebra and a suitable
graded Frobenius algebra ${\cal F}$ \cite{Gaberdiel:1997ia}.

To this end we take ${\cal F}\equiv{\rm Mat}_{2}(\mathbb C)$ 
to be spanned by $(I,J=1,2)$ \cite{Boulanger:2015kfa,Bonezzi:2016ttk}
\be
e_{IJ}=\left[
 \begin{array}{cc}
 e & f \\
 \tilde f & \tilde e \\
 \end{array}
 \right]\ ,\qquad e_{IJ}e_{KL}=\delta_{JK}e_{IL}\;.
\ee
We then define the superconnection $\X$, $sp(2)$ gauge generators $\K_{ij}$, 
and nilpotent differential ${\bf q}$, respectively, by
\be
\X:=A\,e+\wt A\,\tilde e+B\,f-\wt B\,\tilde f\;,\quad
\K_{ij}:=(e+\tilde e)K^{(Y)}_{ij}\;,\quad  \q:=\,(e+\tilde e)d\;,
\ee
introduce the $3$-grading ${\rm deg}_{{\cal F}}(\tilde f,e,
\tilde e,f)=(-1,0,0,1,)$, and use Koszul signs governed
by the total degree given by the sum of form degree 
and ${\rm deg}_{\cal F}$; we note that ${\bf q}$ has total 
degree given by~$1$, while $\X$ has total degree given by $1$ prior 
to duality extension, and in $\{1,3,\dots\}$ after duality extension. 
In terms of these requisites, the equations of motion and gauge conditions
can be written on the desired format as
\be
\q\X+\X\star\X=0\;,\qquad [\K_{ij}, \X]_\star=0\;,\label{4.26}
\ee
and the factorization of the $sp(2)$ ideal amounts to the shift symmetries
\be \delta \X= \K_{ij}\star \boldsymbol{\alpha}^{ij}\ .\label{4.27}\ee
%

\subsection{Comments on action and quantum corrections}

We propose to make the equations of motion \eqref{4.26} (including
the $sp(2)$ gauge condition) variational by taking
the spacetime manifold to be part of the boundary of an open
manifold ${\cal X}$, extending $\X$ to a master field $\widehat \X$
that depends on a set of ghost $(B^{ij},C_{ij})$ variables
obeying
\be \{ B^{ij},C_{kl}\}=\delta_{(k}^i \delta_{l)}^i\ ,\ee
and introducing a master field $\widehat\P$ that vanishes at 
$\partial {\cal X}\times{\cal Z}$.
The Koszul signs are governed by the total degree 
given by the sum of the form degree, degree in ${\cal F}$
and ghost number.
The total degree of $\widehat\X$ lies in $\{1,3,\dots,2p-1\}$,
where ${\rm dim}({\cal X})=2p-3$ or $2p-4$, subject to the
condition that the sum of form degree and degree on ${\cal F}$ 
is non-negative.
The total degree of $\widehat \P$ lies in $\{1,3,\dots,2p-1\}$
if ${\rm dim}({\cal X})=2p-3$ or in $\{0,3,\dots,2p-2\}$
if ${\rm dim}({\cal X})=2p-4$, again subject to the
condition that the sum of form degree and degree on ${\cal F}$ 
is non-negative.

Defining the BRST operator
\be \Q = C^{ij}\K_{ij} -2i  B_i{}^j C_j{}^k C_k{}^i ~, 
\quad \Q^2=0~, \ee
and the covariant derivative
\be \D = \q + {\rm ad}_{\Q} ~, \quad  \D^2=0~, \quad \q\Q=0 ~,  \ee
the flatness condition
\be \D \widehat\X + \widehat\X \star \widehat\X=0\ ,\quad \mbox{at $\partial{\cal X}$}\ ,\label{flat} \ee
follows from the variational principle applied to 
\begin{align}
\label{Sodd}
\mbox{dim(${\cal X}$) odd}\ :\quad S&= \int_{{\cal X} \times {\cal Z}}
{\rm Tr}_{\mathcal A}{\rm Tr}_{\cal F}
{\rm Tr}_{\cal G} \Big(\widehat\P\star(\D\widehat\X + \widehat\X\star\widehat\X) 
+ \tfrac{1}{3} \widehat\P\star\widehat\P\star\widehat\P \Big)~,\\
\label{Seven}
\mbox{dim(${\cal X}$) even}\ :\quad S&= \int_{{\cal X} \times {\cal Z}} {\rm Tr}_{\mathcal A}{\rm Tr}_{\cal F}
{\rm Tr}_{\cal G}\Big(\widehat\P\star(\D\widehat\X + \widehat\X\star\widehat\X) 
+ \tfrac{1}{2} \widehat\P\star\widehat\P \Big)~,
\end{align}
treating ${\cal Z}$ as a closed manifold, and 
where ${\rm Tr}_{\mathcal A}$ denotes the (cyclic) trace operation
over the extended Weyl algebra $\mathcal A$ generated
by polynomials in $Y$, $\kappa_y$ and $k$ (constructed as in
\cite{Boulanger:2015kfa,Bonezzi:2016ttk}); ${\rm Tr}_{\cal F}$
is the standard trace operation on ${\cal F}\equiv{\rm Mat}_{2}$; and  
${\rm Tr}_{\cal G}$
is the standard trace over the Clifford algebra ${\cal G}$ 
generated by the ghosts.
With these definitions, the kinetic term is based on a 
non-degenerate bilinear form. 
Thus, the proposal is that Eqs. \eqref{4.26} and \eqref{4.27}
describe the BRST cohomology contained in \eqref{flat}.

As for boundary conditions, we assume that ${\cal X}\times {\cal Z}$ 
is a compact manifold that contain subregions ${\cal X}'\times{\cal Z}$,
with ${\cal X}'$ corresponding to conformal boundaries, where a 
subset of the master field components are allowed to blow up;
in particular, treating ${\cal Z}$ as a compact manifold
with non-trivial cycles affects the degrees of freedom
that are local on $\partial{\cal X}$, as already commented on above.
The homogenous Dirichlet boundary condition on $\widehat{\P}$ 
does not follow from the classical
variational principle; instead it follows from the requirement 
that the field theory BRST operator is a smooth functional differential 
of a topological field theory \cite{AKSZ,Boulanger:2012bj}. 
The latter property is preserved under the addition of topological
invariants to $\partial {\cal X}\times {\cal Z}$.
If these contain components of $\widehat{\X}$ in sufficiently high form degree, then
they may receive quantum corrections from the $\widehat{\P}^{\star 2}$ 
and $\widehat{\P}^{\star 3}$ vertices.
The topological invariants may thus be non-trivial on-shell, thereby
providing boundary micro-state observables appearing in the
boundary partition function (as $\widehat\X$ is left free to fluctuate
at $\partial{\cal X}\times {\cal Z}$); in addition, if the 
expectation values in $\widehat\X$ at $\partial{\cal X}\times{\cal Z}$ 
(due to non-trivial cycles and including the zero-form initial data) 
source forms in $\widehat\X$ in higher degrees, then the resulting
boundary partition function may contain non-trivial bulk quantum
corrections.
This suggests that the standard (duality unextended) Chern classes, 
which only contain one-forms from $A$ and $\widetilde A$, 
correspond to free conformal theories, while their duality extensions,
which contain higher forms from $A$ and $\widetilde A$, 
correspond to non-trivial conformal field theories.

\section{Conclusions}

In this work, we have first presented an alternative to
Vasiliev's on-shell formulation of the Type A model in 
general spacetime dimensions, using the same field content
but a different $sp(2)$ gauge symmetry with field 
independent generators.
We have argued that this model propagates the same degrees
of freedom as Vasiliev's original equations, and we have
provided evidence that the two models are perturbatively equivalent.
Drawing on the field independence of the $sp(2)$ generators of the new model, we have then extended its equations of motion by a dynamical two-form. This extension requires two connection one-forms, 
gauging the separate left- and right-actions of a complexified higher spin 
algebra, and a zero- and two-form in opposite (real) bi-fundamental 
representations. 
Finally, we have proposed that the latter set of equations 
describes the BRST cohomology of a system that descends from a
variational principle, that is obtained by further extension by 
first-quantized ghosts and an internal graded Frobenius algebra.
If this proposal holds true, then these extensions permit the packaging of the equations 
of motion and the $sp(2)$ gauge conditions, respectively, into a
flatness condition and a set of gauge transformations
for a single odd superconnetion $\widehat\X$.
The action also requires the introduction of a supermomentum 
$\widehat\P$ that may quantum deform certain observables, that
may be of importance in taking the correspondence between 
topological open strings and conformal fields beyond the
current agreement at the level of conformal particles and
free fields \cite{Engquist:2005yt,Colombo:2010fu,Colombo:2012jx,Didenko:2012tv,Bonezzi:2017vha}.

Although the extension with dynamical two-form does not retain 
manifest Lorentz covariance, it is nevertheless suitable for 
potential extensions of higher spin gravity to more general 
non-commutative manifolds. 
Indeed, the extension by the two-form provides a link to  
topological open string fields theory, which is 
the natural framework for deforming non-commutative geometries.

We have deferred a number of technical aspects for future work: 
First of all, it remains to map 
linearized states in lowest weight spaces (particle-like solutions) 
in $\Phi$ to Fronsdal fields in $W_\mu$ by finding a suitable
gauge function; for related supporting results for the four-dimensional 
twistor formulation, see \cite{Iazeolla:2012nf,Sundell2016,Carlo2017}.
Furthermore, in order to establish whether the old and the new Type A 
models are perturbatively equivalent; the first step is to examine whether 
$K_{ij}^{({\rm diag})}$ and $K_{ij}^{(L)}$ agree in Vasiliev-Fronsdal gauge
at first sub-leading order.

As for the formulation in terms of the superconnection $\X$, 
the topology and the boundary conditions of ${\cal X}\times
{\cal Z}$ need to be examined.
In particular, ${\cal Z}$ needs to 
contain a non-trivial two-cycle in order for the dynamical two-form to
contain the original closed and central element as a non-trivial
vacuum expectation value. In this case the alternative Type A master fields
arise as a consistent truncation of $\X$; if so, however,
the dynamical two-form leads to
new local degrees of freedom in spacetime, whose holographic 
interpretation remains to be given; for related issues in the case
of the four-dimensional twistor theory, see \cite{Boulanger:2015kfa,Bonezzi:2016ttk}.

Our proposal for an action, producing the $sp(2)$ condition as well 
from a variational principle, relies on the claim made in Section 4.3 
concerning the BRST cohomology contained in the flat superconnection
$\widehat \X$ (obtained by extension by first-quantized $sp(2)$ ghosts).
In the aforementioned action principle, the $sp(2)$ generators are fixed 
given operators. In this context, it would be interesting to treat them 
as new fluctuating degrees of freedom 
\cite{Bars1997,Bonezzi2010,Bonezzi2014,Bekaert2017}
of an enlarged string field.

Concerning the basic physical motivation behind our work, namely that from the recent gathering of results concerning
 the nature of the Noether procedure, it appears that the formulation
 of higher spin gravity in terms of Fronsdal fields leads to a perturbatively defined quantum effective action making 
 sense in asymptotically maximally symmetric spacetimes, whereas the 
 topological open string field theory formulation provides perturbative
 expansions around more general backgrounds.
 In addition, the latter formulation leads to the notion of star product 
 locality, whereby the classical action is built from data obtained
 from disc amplitudes, thus replacing the more subtle notion of spacetime 
 quasi-(non)locality that needs to be adopted following the standard Noether 
 approach.

Finally, we remark that the alternative $sp(2)$ gauging for the Type A
model presented in this work has a direct generalization to the Type B 
model based on $osp(1|2)$ gauging, whose conformal field theory dual
expanded around the anti-de Sitter vacuum consists of free fermions;
we hope to present this model in more detail in a forthcoming work.

\paragraph{Acknowledgements:} 
We have benefited from conversations with N. Boulanger, C. Iazeolla, 
E. Sezgin and A. Waldron.
C.A. is a Universidad Andres Bello (UNAB) Ph.D. Scholarship holder, 
and his work is partially supported by Direcci\'on General de
Investigaci\'on (DGI-UNAB).
R. B. would like to thank the hospitality of UNAB.
The work of R.B. is supported by a PDR “Gravity and
extensions” from the F.R.S.-FNRS (Belgium).
P. S. would like to thank the hospitality of the University of Bologna.
The work of P.S. is supported by Fondecyt Regular grants N$^{\rm o}$ 1140296 and N$^{\rm o}$ 1151107 and Conicyt grant DPI 2014-0115.


\end{document}